\documentclass[aps,prl,showpacs,twocolumn,superscriptaddress,floatfix]{revtex4}

\usepackage{graphicx}
\usepackage{dcolumn}
\usepackage{bm}
\usepackage{amsmath}
\usepackage{amssymb}

\begin{document}

\renewcommand{\vec}[1]{\mbox{\boldmath $#1$}}


\title{Emergence of highly degenerate excited states in frustrated magnet MgCr$_2$O$_4$}


\author{K. Tomiyasu}
\email[Electronic address: ]{tomiyasu@m.tohoku.ac.jp}
\affiliation{Department of Physics, Tohoku University, Aoba, Sendai 980-8578, Japan}
\author{T. Yokobori}
\affiliation{Department of Physics and Mathematics, Aoyama-Gakuin University, Sagamihara 229-8558, Japan}
\author{Y. Kousaka}
\affiliation{Department of Physics and Mathematics, Aoyama-Gakuin University, Sagamihara 229-8558, Japan}
\author{R. I. Bewley}
\affiliation{ISIS Facility, Rutherford Appleton Laboratory, Chilton, Didcot, OX11 0QX, UK}
\author{T. Guidi}
\affiliation{ISIS Facility, Rutherford Appleton Laboratory, Chilton, Didcot, OX11 0QX, UK}
\author{T. Watanabe}
\affiliation{Department of Physics, College of Science and Technology (CST), Nihon University, Chiyoda, Tokyo 101-8308, Japan}
\author{J. Akimitsu}
\affiliation{Department of Physics and Mathematics, Aoyama-Gakuin University}
\author{K. Yamada}
\affiliation{Institute of Material and Structure Science, High Energy Accelerator Research Organization, Oho, Tsukuba, Ibaraki 305-0801, Japan}


\date{\today}

\begin{abstract}
High degeneracy in ground states leads to the generation of exotic zero-energy modes, a representative example of which is the formation of molecular spin liquid-like fluctuations in a frustrated magnet. Here we present single-crystal inelastic neutron scattering results for the frustrated magnet MgCr$_2$O$_4$, which show that a common set of finite-energy molecular spin excitation modes is sustained in both the liquid-like paramagnetic phase and a magnetically ordered phase with an extremely complex structure. Based on this finding, we propose the concept of high degeneracy in excited states, which promotes local resonant elementary excitations. This concept is expected to have ramifications on our understanding of excitations in many complex systems, including not only spin but also atomic liquids, complex order systems, and amorphous systems.
\end{abstract}

\pacs{75.25.Dk, 75.40.Gb, 78.70.Nx, 63.50.-x, 36.10.-k}

\maketitle

%
%

The concept of elementary excitations, or quasiparticles, constitutes the basis of modern condensed matter physics~\cite{Landau_1981}. Intricate interactions among a large number of particles, like in liquids, can be successfully treated as a collection of independent quasiparticles; examples include magnons, phonons, and rotons arising from helium superfluids~\cite{Feynman_1954}. In addition, magnetic pseudo-monopoles, which have been long sought after in high-energy physics, were recently observed as elementary excitations in a spin liquid-like paramagnetic phase in a highly frustrated magnet called a spin ice~\cite{Gingras_2009}. Thus, studying elementary excitations in complex systems like liquids could have broad implications across many fields of physics.

\begin{figure}[htbp]
\begin{center}
\includegraphics[width=3.0in, keepaspectratio]{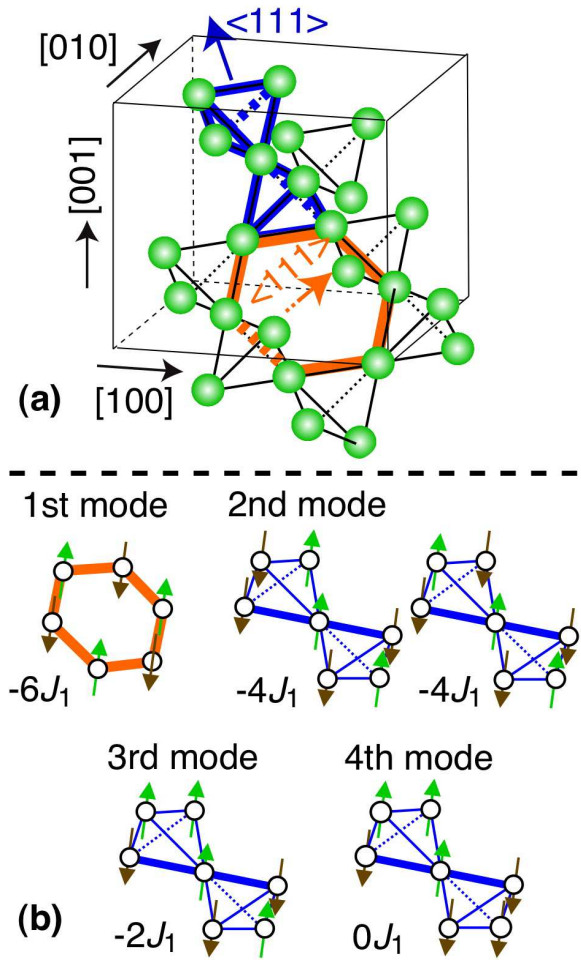}
\end{center}
\caption{\label{fig:models} (Color online)
(a) Pyrochlore lattice. The bold lines mark examples of positions of the hexamer and heptamers shown in (b) among equivalent positions.
(b) Snapshots of spin molecular models. The arrows indicate spins, which dynamically fluctuate in arbitrary directions in keeping with the relative correlations.
The total exchange interaction energies in the molecules are also shown, where $J_{1}$ is the magnitude of the nearest-neighbor exchange interaction energy.
}
\end{figure}

\begin{figure*}[htbp]
\begin{center}
\includegraphics[width=0.90\linewidth, keepaspectratio]{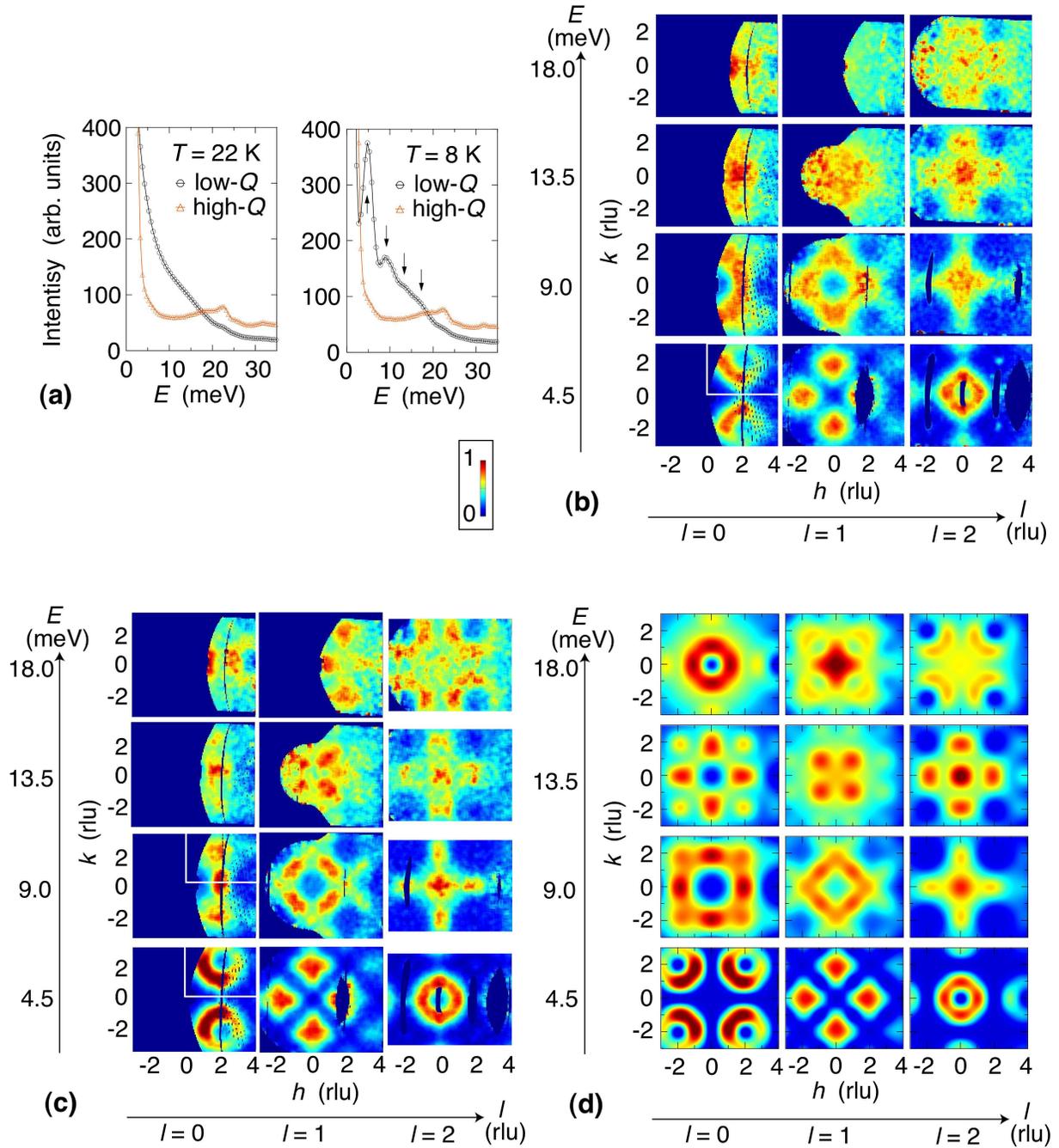}
\end{center}
\caption{\label{fig:data} (Color online)
(a) $E$ dependence of $\vec{Q}$-integrated intensity measured in a paramagnetic phase (left panel) and magnetically ordered phase (right panel). The integration ranges are $0<h<2$, $0<k<2$, and $0<l<2$ reciprocal lattice units (rlu) in the low-$Q$ data and $0<h<2$, $5<k<7$, and $0<l<2$ rlu in the high-$Q$ data. 
The arrows indicate the magnetic peaks and shoulders. The errors are smaller than the symbols used. The lines are guides to the eye.
(b)(c) Measured inelastic neutron scattering intensity distributions in $(h,k,l,E)$ space in a paramagnetic phase (22 K) and magnetically ordered phase (8 K), respectively. Each set of three-abreast panels corresponds to the data for the given $E$ measured in the $hk$0, $hk$1, and $hk$2 planes. The intensities were integrated over ranges of $\pm0.2$ rlu and $\pm1.0$ meV around the given $l$ and $E$, respectively. The vertical tone indicates the linear scale intensity in arbitrary units in (b)--(d). The three first quadrants, boxed by white lines in the $hk$0 planes, show areas previously reported~\cite{Tomiyasu_2008}.
(d) One-to-one correspondence between calculated patterns as identified by the molecular models shown in Fig.~\ref{fig:models}(b).
}
\end{figure*}

Highly frustrated magnets are promising sources for exotic spin liquid-like states. This is because in frustrated magnets, not all classical-spin pairs can be arranged antiferromagnetically on a triangular or tetrahedral lattice, which gives rise to an inherent macroscopic degeneracy in ground states~\cite{Wannier_1950,Anderson_1956}. Therefore, in a low-temperature paramagnetic phase, frustration suppresses magnetic ordering and promotes spin liquid-like fluctuations (zero-energy modes), which are accompanied by short-range spatial correlations that lower system entropy, as required by the third law of thermodynamics. A representative example of this phenomenon is the formation of molecular spin liquid-like fluctuations, where a spin molecule refers to a spin cluster that is spatially confined within a geometrical region, such as an atomic molecule.
For example, in the spinel antiferromagnet MgCr$_2$O$_4$ ($T_{\rm N}=13$ K), the magnetic ions Cr$^{3+}$ ($(t_{2g})^3$, spin $S=3/2$) form a corner-sharing tetrahedral lattice called a pyrochlore lattice, as shown in Fig.~\ref{fig:models}(a), and the paramagnetic phase exhibits antiferromagnetic spin hexamers~\cite{Tomiyasu_2008}, as shown as the first mode in Fig.~\ref{fig:models}(b). The hexamers correspond to a characteristic neutron scattering intensity pattern that is widely spread along the Brillouin zone boundary~\cite{Tomiyasu_2008}, as shown in the bottom-left corner panel in Fig.~\ref{fig:data}(b), which is reproduced by classical-spin Monte Carlo simulations~\cite{Conlon_2010}.

%

Notably, there is little available information on {\it finite-energy} elementary excitations in this molecular spin liquid-like paramagnetic phase; in fact, there is no experimental report in existence. This is probably because zero-energy modes are one of the most important phenomena directly resulting from frustration. Indeed, many studies on frustrated magnets in the paramagnetic phase were focused solely on the zero-energy modes~\cite{Tomiyasu_2008,Lee_2002,Chung_2005,Tomiyasu_2011a,Yasui_2002}.

By contrast, in the magnetically ordered phase in several spinel antiferromagnets, where frustration was assumed to be relieved by a lattice distortion, molecular spin and spin-orbit resonances were recently discovered to exist as non-dispersive gapped elementary excitation modes~\cite{Tomiyasu_2008,Tomiyasu_2011a,Tomiyasu_2011b}. An antiferromagnetic phase in MgCr$_2$O$_4$ exhibits a complex tetragonal spin-lattice order that is essentially equivalent to that in ZnCr$_2$O$_4$~\cite{Shaked_1970,Ji_2009}, and the hexamer mode and a heptamer mode (the first and second modes in Fig.~\ref{fig:models}(b)) are observed at 4.5 meV and 9.0 meV, respectively~\cite{Tomiyasu_2008}. Interestingly, the heptamer-type zero-energy mode is also observed in a paramagnetic phase in another frustrated pyrochlore magnet Tb$_2$Ti$_2$O$_7$~\cite{Yasui_2002}, and a ground state composed of the same structural units is observed in frustrated spinel magnet AlV$_2$O$_4$ with charge degree of freedom~\cite{Horibe_2006}. Thus, the magnetically ordered phase does incorporate some sort of frustration effect in its excitations.

In this study, we explored finite-energy excitations in the paramagnetic phase and clarified their relation with those in the magnetically ordered phase through a comprehensive study of spin excitations distributed over wide momentum ($\vec{Q}$) and energy ($E$) ranges both in a paramagnetic and magnetically ordered phase in MgCr$_2$O$_4$ by inelastic neutron scattering. The overall study of spin excitations was possible because of the combination of a large single-crystal assembly, an advanced time-of-flight spectrometer with large-solid-angle detectors, and sophisticated software to handle huge data sets in four-dimensional space.

%
%
Inelastic neutron scattering experiments were performed using the direct geometry chopper spectrometer MERLIN at the ISIS (UK) spallation neutron source~\cite{Bewley_2006}. The detector coverage is as large as $-45^{\circ}$ to $135^{\circ}$ in the horizontal plane and $\pm30^{\circ}$ in the vertical direction. The incident energy ($E_i$) was fixed at 50 meV with a chopper speed of 400 Hz. The energy resolution under elastic conditions was approximately 5{\%} of $E_i$.
Single-crystal rods of MgCr$_2$O$_4$ were grown by a floating zone method. Details of the crystal growth are summarized in Ref.~\cite{Kousaka_2011}. The rod was about 4 mm in diameter and 40 mm long. Six co-aligned single crystals were fixed by thin aluminum plates and inserted in a closed-cycle $^4$He refrigerator with $^4$He exchange gas.
Since the spin system is three-dimensional, as shown in Fig.~\ref{fig:models}(a), the data were recorded while rotating the crystal in 1$^{\circ}$ steps about the vertical axis, which is unlike experiments on two-dimensional and one-dimensional systems with a fixed crystal angle ($\omega$) ~\cite{Tranquada_2004,Arai_1996}. The huge combined data sets were handled by the HORACE software of ISIS~\cite{Perring_URL}.

%
%

%
%

Figure~\ref{fig:data}(a) shows the $E$ spectra of $\vec{Q}$-integrated intensity, measured in a paramagnetic phase (22 K) and magnetically ordered phase (8 K). In neutron scattering, the magnetic scattering intensity decreases by a magnetic form factor while the phonon intensity increases in proportion to $Q^2$ with increasing $Q$~\cite{Marshall_1971}.
Since below $\sim20$ meV, the low-$Q$ intensity is higher than the high-$Q$ intensity, as shown in Fig.~\ref{fig:data}(a), the signals in this low-energy region are magnetic in origin.
As shown in the left panel, a single quasielastic mode seems to exist above $T_{\rm N}$. In contrast, as shown in the right panel, peaks around 4.5 meV and 9.0 meV and tiny shoulders around 13.5 meV and 18.0 meV are observed below $T_{\rm N}$, which are of equal energy interval. 

Despite the appearance of a single mode, we investigated the scattering intensity distributions in a $\vec{Q}$ space sliced at the four characteristic $E$'s not only below but also above $T_{\rm N}$, as shown in Figs.~\ref{fig:data}(b) and \ref{fig:data}(c), respectively. Above $T_{\rm N}$ (Fig.~\ref{fig:data}(b)), different patterns spreading over several rlu appear at 9.0 meV, 13.5 meV, and 18.0 meV in addition to the zero-energy hexamer patterns at 4.5 meV, indicating the existence of other extremely short-range correlation modes. Further, as shown in Fig.~\ref{fig:data}(c), all the modes are sustained below $T_{\rm N}$: the 4.5-meV patterns are almost identical. Each 9.0-meV pattern above $T_{\rm N}$ is regarded as a superposition of the 4.5-meV pattern and the heptamer 9.0-meV pattern below $T_{\rm N}$ and, thus, the heptamer mode certainly exists both below and above $T_{\rm N}$, and the remaining 13.5-meV and 18.0-meV patterns above $T_{\rm N}$ are smeared compared with those below $T_{\rm N}$ but are essentially the same. Here, the 4.5-meV and 9.0-meV patterns below $T_{\rm N}$ are consistent with the previous report~\cite{Tomiyasu_2008}. Thus, in the paramagnetic phase with the zero-energy mode, we found multiple finite-energy short-range correlation modes, all of which are also in common with the magnetically ordered phase. The appearance of a single mode is due to superposition of all the modes that are broader in $E$ compared to the ordered phase. 

%
%
Next, to extract information on the spatial correlations of these modes, we tried to reproduce the experimental $\vec{Q}$ patterns as two-body correlation functions of classical spins, like in Refs.~\cite{Tomiyasu_2008,Lee_2002}. The Watson--Freeman magnetic form factor of Cr$^{3+}$~\cite{Watson_1961} and the orientation average over the equivalent directions were also taken into account. Through trial and error, heptamer variation models were also found for the third and fourth modes, as shown in Fig.~\ref{fig:models}(b), and the corresponding calculated patterns (Fig.~\ref{fig:data}(d)) are in agreement with those of Figs.~\ref{fig:data}(b) and \ref{fig:data}(c). 
Further, the total exchange interaction energy evaluated from the numbers of antiferromagnetic and ferromagnetic nearest-neighbor bonds is of equal interval as follows, which is also in agreement with experiments: the energy for each molecule can be evaluated to be $-6J_1$, $-4J_1$, $-2J_1$, and $0J_1$ with an equal interval of $2J_1$, where $J_{1}$ denotes the magnitude of an antiferromagnetic first-neighbor exchange interaction energy; this first-neighbor exchange interaction has been reported to be predominant over other exchange interactions in band calculations~\cite{Yaresko_2008}. 
However, it is uncertain whether these are a unique solution.

%
%
To summarize the aforementioned results, the present experiments found a common hierarchical set of spin excitations ranging from the zero-energy mode to the finite-energy modes above and below $T_{\rm N}$, which can be described by the hexamer and heptamer variation models. Now we turn to the significance of this finding. First, existence both above and below $T_{\rm N}$ means that the molecular spin excitations are neither caused by magnetic ordering nor likely to be normal spin waves arising from the magnetic order. In addition, the spin-lattice order is tetragonal~\cite{Shaked_1970,Ji_2009}, whereas the hexamers and heptamers are trigonal, as shown in Fig.~\ref{fig:models}(a); they are also different in symmetry. This difference in symmetry has been very recently detected by ultrasound measurements in the paramagnetic phase: there are tetragonal spin excitations/fluctuations below $\sim3.4$ meV originating from spin Jahn--Teller coupling with the tetragonal spin-lattice order and two types of trigonal spin excitations around $\sim3.4$ meV and in the higher-energy region, corresponding to the hexamer and the heptamer variations, respectively~\cite{Watanabe_2012}.

Second, we focus on the finding that the finite-energy modes are spatially confined, as is the zero-energy mode. The spatially confined zero-energy mode is recognized as being the direct result of high degeneracy in the ground states, which restricts the formation of a normal wavelike magnetic order that is inevitably longer than the wavelength of the propagation vector. In analogy with this, the spatially confined finite-energy modes suggest {\it high degeneracy in excited states}, or the expanded concept of {\it frustration in excitations}. Indeed, this concept seems to be realized in the magnetically ordered phase, which exhibits a complex magnetic structure with multiple propagation vectors and 32 magnetic ions in its unit cell~\cite{Shaked_1970,Ji_2009}, implying that 32 spin wave modes are squeezed in an energy region. This situation is most probably a clearer example of the term ``frustration effect in excitations" mentioned in the previous report~\cite{Tomiyasu_2008}. Similarly, in the paramagnetic phase without a magnetic order, since the concept of a unit cell is broken down or a unit cell with the larger numbers of magnetic ions can be defined, further higher degeneracy is expected in the spin excited states. Thus, one can conclude that although the molecular modes might have originally been normal spin waves, the expanded frustration transforms them beyond recognition.

Last, we discuss the ramifications of this expanded concept. The concept is realized when a unit cell consists of many components. However, in contrast, from a mean-field theory for pyrochlore systems with only four magnetic sublattices, it was reported that only ground states are highly degenerated and that this is not the case for excited states~\cite{Reimers_1991}. This, in turn, suggests that the expanded concept can be also applied to atomic liquids and amorphous systems, which are similar to spin frustrated systems as follows: a spin-frustrated system exhibits, for example, a spin glass state because of the presence of trace impurities~\cite{Ratcliff_2002}, which is comparable to a supercooled liquid that undergoes a glass transition upon experiencing a small impact. Further, supercooled liquids and glasses exhibit similar resonance-like short-range vibrational modes called boson peaks. Interestingly, a recent X-ray spectroscopic study revealed that boson peaks are originally identical to, but are then severely transformed from, lattice waves propagating in a crystal (phonons), and also suggested that the $\vec{Q}$ correlations are distributed over several rlu near the pseudo Brillouin zone boundary~\cite{Chumakov_2011}.

%
%
In summary, we studied spin excitations over a wide $(\vec{Q},E)$ space in the frustrated spinel magnet MgCr$_2$O$_4$ by single-crystal time-of-flight inelastic neutron scattering. A set of molecular spin excitation modes was commonly observed above and below $T_{\rm N}$. This observation leads us to the concept of frustration in excitations, which is applicable both above and below $T_{\rm N}$, and probably also in atomic liquids, complex ordered systems, and amorphous systems.

\acknowledgments
We thank Professors T. Masuda and T. J. Sato for the fruitful discussions and Mr. M. Onodera for providing assistance at Tohoku University. This study was financially supported by Grants-in-Aid for Young Scientists (B) (22740209), Priority Areas (22014001), and Scientific Researches (S) (21224008) and (A) (22244039) from the MEXT of Japan.

\bibliography{MgCr2O4_4_PRL}

\end{document}